\documentclass[aps,prd,showpacs,preprint,preprintnumbers,showpacs,showkeys,superscriptaddress,amsmath,amssymb,nofootinbib]{revtex4-1}
\usepackage{amsmath}
\usepackage{latexsym}
\usepackage{graphicx}
\usepackage{pstricks,pst-coil}
\usepackage{pst-all}
\usepackage{color}
\usepackage{latexsym}
\usepackage{amsmath}
\usepackage{amssymb}
\usepackage{eufrak}
\usepackage{euscript}
\usepackage{graphics}
\usepackage{picture}
\begin{document}

\title{\Large{A Dark Hidden-Sector of Dirac fermions at the GeV scale}}

\author{M. J. Neves}\email{mariojr@ufrrj.br}
\affiliation{Department of Physics, Federal Rural University of Rio de Janeiro,
BR 465-07, Serop\'edica, RJ, Brazil, Zip code 23890-971}

\author{J. A. Hela\"yel-Neto}\email{helayel@cbpf.br}
\affiliation{Brazilian Centre for Research in Physics, Street Dr. Xavier Sigaud
150, Urca, Rio de Janeiro, Brazil, Zip code 22290-180}



\begin{abstract}

Our contribution sets out to investigate the GeV-scale phenomenology of a model based on an $SU_{L}(2) \times U_{Y}(1) \times U(1)_{X}$-gauge symmetry. The model accommodates, as a consequence of the symmetry-breaking pattern, a light gauge boson at the GeV-scale or below, allowing then to set up a new low energy physics. The fermion sector includes an exotic candidate to Dark Matter (DM) and the whole fermionic field content yields, as it is mandatory, the cancellation of the chiral anomaly.
The light gauge boson and the exotic fermion are proposed as candidates to the so-called dark sector of the model that we try
to describe in this contribution. The low-energy spectrum exhibits a hidden scalar field with mass fixed at the GeV level in the Higgs sector. We calculate the DM relic density associated with the dark fermion candidate by considering two annihilation processes in which the light gauge boson works as a portal for the DM detection. The observed relic density points to a mass of $3.4$ GeV for the DM fermion, while the light gauge boson mass is $8.0$ GeV. We also investigate the direct detection of DM in the low energy regime and obtain the region of parameters allowed by recent discussions on DM limits. Finally, the correction to the muon magnetic dipole moment is calculated considering the gauge boson mass at the GeV scale or below.

%
\end{abstract}
%
%
%
%

\maketitle

\pagestyle{myheadings}
\markright{A Dark Hidden-Sector of Dirac Fermions at the GeV Scale}

%
%
%

%
\section{Introduction}

The scenario of particle physics at the GeV scale (1-10 GeV) and below has attracted the attention of experimental research groups \cite{BaBarPRL2017,NA64collaboration1,NA64collaboration2,BaBarPRD2016}. The search for a single light photon at $53 \, \mbox{fb}^{-1}$,
in the $e^{+} \, e^{-}$ CM-collision with the production of a spin-1 particle, $A'$, referred to the literature as a dark photon, is associated with the process $e^{+} \, e^{-} \, \rightarrow \, \gamma \, A'$ \cite{BaBarPRL2017}. Immediately, the dark photon decays into invisible fermion matter $A' \, \rightarrow \, \bar{\zeta} \, \zeta$, and the $\zeta$-fermion can be a candidate to be included in the Dark Matter (DM) content. The dominant decay mode for the lowest-mass of the $\zeta$-state fixes the condition $M_{A'} \, > \, 2 \, M_{\zeta}$, considering that the $M_{A'}$-dark photon mass is bounded from above by $M_{A'} \leq 8 \, \mbox{GeV}$. The $A'$-gauge boson is introduced in the Standard Model (SM) framework as part of a new Abelian sector that mixes kinetically with the usual electromagnetic photon. Therefore, this scenario based on $A'$ with an invisible sector may work as a possible portal for a DM detection. The introduction of the $A'$-light gauge boson has also been considered in recent models to explain the muon $(g-2)_{\mu}$ muon anomaly \cite{AstJean2003,Knodlseder2003,Pamela,Ackermann,Berezhiani,Bennet}.
In the context of a particular Nuclear Physics experiment, anomalies in the decay of the excited state of $8\, \mbox{Be}^{\ast}$ to its ground state have suggested the existence of a new neutral boson, dubbed $X$-boson, through the decay mode $8 \, \mbox{Be}^{\ast} \rightarrow 8 \, \mbox{Be} + X$\cite{KrasPRL2016,FengPRL2016,JFengPRD2017}. The $X$-boson immediately decays into an electron-positron pair $X \rightarrow e^{+} \, + \, e^{-}$. It is a vector-type (spin-1) neutral particle, with mass around $m_{X}=17 \, \mbox{MeV}$ that mixes kinetically with the SM photon. Its origin could, in principle, be traced back to an extra gauge symmetry, $U(1)_{X}$, in addition to the symmetries of the SM.
Thereby, the $X$-boson has a similar phenomenology to the $A'$-dark photon, with a kinetic mixing of order $10^{-3}$, with the EM photon.
A considerable number of models has been proposed to explain the $X$-Boson phenomenology in the literature \cite{GuHe2016,MJNeves2017Annalen,KahnJHEP2017,BilmisPRD2015,KitaharaPRD2017,KozaczukPRD2017,Moretti2019}.

The well-known $(B-L)$-model describes an extra gauge boson is based on the gauge symmetry $SU_{L}(2) \times
U_{Y}(1) \times U(1)_{B-L}$, which includes a $U(1)_{B-L}$-extra group in the SM, where the charge is defined by
the baryon number (B) minus the leptonic number (L)
\cite{LangackerRMP2009,Kanemura2011,CaiPLB2018,Patra,Yaguna2015,HanEPJC2018,NandaPRD2017,MandalPRD,Nobu2010}.
The gauge sector has just one extra boson, while the Higgs is extended to include a doublet and a singlet, or bi-doublets of scalars fields.
We should keep in mind that the introduction of an extra Higgs may be needed to explain the gauge boson mass at the $\mbox{GeV}$-scale, or below. These masses are associated with a new range of VEVs of the extra Higgs scalar. The fermion sector is enlarged to ensure anomaly cancellation; for example, the introduction of right-neutrinos components and exotic fermions singlets that could be candidates to a DM content.


In the present paper, we re-assess the $SU_{L}(2) \times U_{Y}(1) \times U(1)_{X}$ model spontaneous symmetry breaking
(SSB) pattern is takes places in a scenario of energy at the GeV scale to include a light gauge boson $(A')$ and an exotic
Dirac fermion, that we call $\zeta$-fermion, and it can be a candidate to DM.
We calculate the relic density of DM associated with the DM fermion, and we study its phenomenology such that the light $A'$-gauge boson
can connect the SM fermions with the DM fermion candidate. In particular, two annihilation processes are important for the DM detection : $ \zeta \, \bar{\zeta} \, \rightarrow \, A' \, \rightarrow \, f \, \bar{f}$ and $\zeta \, \bar{\zeta} \, \rightarrow \, A' \, A'$. The first case yields the decay processes of $A'$ into SM fermions and DM fermion, if it is kinetically possible. Afterwards, we analyse the direct detection of DM and the bounds of XENON1T experiment. We also calculate the contribution of the $A'$-gauge boson to the muon magnetic moment, for the $A'$ mass at the GeV scale and below.

The organization of this paper follows the outline below: in Section II, we present the model content based on an $SU_{L}(2) \times U_{Y}(1) \times U(1)_{X}$-symmetry. In the section III, we obtain the massive eigenstates and correspondent neutral currents.
In Section IV, we calculate the relic density for $\zeta$-fermion DM and we obtain the decay modes of the $A'$-gauge boson.
In Section V, we discuss briefly the direct detection of DM.
%
The section VI addresses to the problem of correcting the muon magnetic dipole moment whenever the model displays a light gauge boson.
Finally, our Concluding Comments are cast in Section VII.
%
%
%

%
\section{The field and particle content of the model}
The light gauge boson framework is introduced through the SSB mechanism
in which the new VEV-scale $(v_{x})$ is at a lower scale with respect to the SM scale, $v_{x} \ll v=246 \, \mbox{GeV}$.
The original $SU_{L}(2) \times U_{Y}(1) \times U(1)_{X}$-symmetry is broken to the final electromagnetic symmetry $U(1)_{em}$.
%
%
%
%
This SSB pattern has the VEV scale, $v_{x}$, that is responsible for the lighter massive gauge boson. It may describe the physics of the dark photon with a mass bound $M_{A'} < 8 \, \mbox{GeV}$ \cite{BaBarPRL2017}, the $X$-boson of mass $17 \, \mbox{MeV}$, or it can also describe the para-photon physics at the $\mbox{Sub-eV}$-scale.
The electric charge of the model remains like the one in the SM :
$Q_{em}=I_{3L}+Y$. The $U(1)_{X}$-sector of $X^{\mu}$ couples kinetically with $Y^{\mu}$ of $U_{Y}(1)$ by means
of a mixing parameter, $\chi$
\begin{equation}\label{Lgauge}
{\cal L}_{gauge}=-\frac{1}{2} \, \mbox{tr}\left(F_{\mu\nu}^{\; 2}\right)
-\frac{1}{4} \, Y_{\mu\nu}^{\; 2}
-\frac{1}{4} \, X_{\mu\nu}^{\; 2}
+\frac{\chi}{2} \, Y_{\mu\nu} \, X^{\mu\nu} \; ,
\end{equation}
%
where it is currently estimated that $\chi \simeq 10^{-3}$ for models that discuss the kinetic mixing of hidden photons in connection with the usual EM photon. The field strengths are defined as, usually, by $F_{\mu\nu}=\partial_{\mu}A_{\nu}-\partial_{\nu}A_{\mu}+ig \, \left[ \, A_{\mu} \, , \, A_{\nu} \, \right]$,  $X_{\mu\nu}=\partial_{\mu}X_{\nu}-\partial_{\nu}X_{\mu}$ and $Y_{\mu\nu}=\partial_{\mu}Y_{\nu}-\partial_{\nu}Y_{\mu}$.

The fermions are minimally coupled to the gauge bosons through the covariant derivative:
%
%
\begin{eqnarray}\label{DmuPsi}
\!\!\!\!\!\!D_{\mu} \psi_{i} = \left( \partial_{\mu} + i \, g \, A_{\mu}^{\, a} \, \frac{\sigma^{a}}{2} + i \, Y_{\psi_{i}} \, g_{y} \, Y_{\mu}
+ i \, X_{\psi_{i}}
\, g_{x} \, X_{\mu} \phantom{\frac{1}{2}} \!\!\!\!\! \right) \psi_{i} \, ,
\end{eqnarray}
where $g$, $g_{y}$ and $g_{x}$ are the dimensionless coupling constants associated with the symmetry groups whereas $Y$ and $X$ are the charges of $U_{Y}(1)$ and $U(1)_{X}$, respectively. The $\psi_{i}$-fields represent any fermion of the model whose content is given below :
%
%
%
%
%
%
\begin{eqnarray}
L_{i} &=&
\left(
\begin{array}{c}
\nu_{e} \\
\ell_{e} \\
\end{array}
\right)_{L} ,
\left(
\begin{array}{c}
\nu_{\mu} \\
\ell_{\mu} \\
\end{array}
\right)_{L},
\left(
\begin{array}{c}
\nu_{\tau} \\
\ell_{\tau} \\
\end{array}
\right)_{L}
\hspace{-0.1cm}
: \left(\underline{{\bf 2}}, -\frac{1}{2}, -1 \right) \, ,
\nonumber \\
\ell_{iR} &=& \left\{ \, e_{R} \, , \, \mu_{R} \, , \, \tau_{R} \, \right\} : \left(\underline{{\bf 1}}, -1, -1 \right) \, ,
\nonumber \\
N_{iR} &=& \left\{ \, N_{eR} \, , \, N_{\mu R} \, , \, N_{\tau R} \, \right\} : \left(\underline{{\bf 1}}, 0, -1 \right) \, ,
\nonumber \\
Q_{iL}&=&
\left(
\begin{array}{c}
u \\
d \\
\end{array}
\right)_{L} ,
\left(
\begin{array}{c}
c \\
s \\
\end{array}
\right)_{L} ,
\left(
\begin{array}{c}
t \\
b \\
\end{array}
\right)_{L}
\hspace{-0.1cm}
: \left(\underline{{\bf 2}}, +\frac{1}{6}, +\frac{1}{3} \right) \, ,
\nonumber \\
%
%
Q_{iR} &=& \left\{ \, u_{R} \, , \, c_{R} \, , \, t_{R} \, \right\} : \left(\underline{{\bf 1}}, +\frac{2}{3}, +\frac{1}{3} \right) \, ,
\nonumber \\
q_{iR} &=& \left\{ \, d_{R} \, , \, s_{R} \, , \, b_{R} \, \right\} : \left(\underline{{\bf 1}}, -\frac{1}{3}, +\frac{1}{3} \right) \, , \,
\zeta_{L(R)}
: \left(\underline{{\bf 1}}, 0, +1 \right) \, .
\end{eqnarray}
The charge values are displayed in Table \ref{Table1}, and the model with a minimal $U(1)_{X}$
is quiral and gravitational anomaly-free. The $\zeta$-fermion is electrically neutral; it just sets the $X$-charge as an example of a hidden charge.
%
%
%
%
\begin{table}
\centering
\begin{tabular}{l l l l l l}
\hline
\mbox{Fields} \quad \quad & \quad \quad $SU_{L}(2)$ \quad & \quad $U(1)_Y$ \quad & \quad $U(1)_X$ \\
\hline
\hline
$L_{i}$ \quad \quad & \quad \quad \quad $\underline{{\bf 2}}$ \quad & \quad $-1/2$ \quad & \quad $-1$  \\
\hline
$\ell_{iR}$ \quad \quad & \quad \quad \quad $\underline{{\bf 1}}$ \quad & \quad $-1$ \quad & \quad $-1$ \\
\hline
$N_{iR}$ \quad \quad & \quad \quad \quad $\underline{{\bf 1}}$ \quad & \quad $0$ \quad & \quad $-1$ \\
\hline
$\zeta_{L(R)}$ \quad \quad & \quad \quad \quad $\underline{{\bf 1}}$ \quad & \quad $0$ \quad & \quad $+1$ \\
\hline
$Q_{iL}$ \quad \quad & \quad \quad \quad $\underline{{\bf 2}}$ \quad & \quad $+1/6$ \quad & \quad $+1/3$  \\
\hline
$Q_{iR}$ \quad \quad & \quad \quad \quad $\underline{{\bf 1}}$ \quad & \quad $+2/3$ \quad & \quad $+1/3$ \\
\hline
$q_{iR}$ \quad \quad & \quad \quad \quad $\underline{{\bf 1}}$ \quad & \quad $-1/3$ \quad & \quad $+1/3$ \\
\hline
$\Phi$ \quad \quad & \quad \quad \quad $\underline{{\bf 2}}$ \quad & \quad $+1/2$ \quad & \quad $0$  \\
\hline
$\Xi$ \quad \quad & \quad \quad \quad $\underline{{\bf 1}}$ \quad & \quad $0$ \quad & \quad $+2$   \\
\hline
\end{tabular}
\caption{The anomaly-free particle content for the $U(1)_{X}$-model with a
$A'$ light gauge boson.} \label{Table1}
\end{table}
The effective Higgs sector is made up by two scalars fields, $\Phi$ and $\Xi$, and its full Lagrangian density reads as follows :
\begin{eqnarray}\label{LHiggs}
{\cal L}_{Higgs} &=&
\left(D_{\mu}\Xi\right)^{\dagger} D^{\mu} \Xi
-\mu_{\Xi}^{\, 2} \, |\Xi|^{2} -\lambda_{\Xi} \, |\Xi|^{4}
+\left(D_{\mu}\Phi\right)^{\dagger} D^{\mu} \Phi
-\mu_{\Phi}^{\, 2} \, |\Phi|^{2}-\lambda_{\Phi} \, |\Phi|^{4}
-\lambda \, |\Xi|^{\, 2} \, |\Phi|^{\, 2}
\nonumber \\
&&
\hspace{-0.5cm}
- \, f_{ij}^{(\ell)} \, \overline{L}_{i} \, \Phi \, \ell_{jR}
- \, f_{ij}^{(d)} \, \overline{Q}_{iL} \, \Phi \, q_{jR}
- \, f_{ij}^{(u)} \, \overline{Q}_{iL} \, \widetilde{\Phi} \, Q_{jR}
- f_{ij}^{(\nu)} \, \overline{L}_{i} \, \widetilde{\Phi} \, N_{j R}
-M_{\zeta} \, \overline{\zeta}_{L} \, \zeta_{R}
+\mbox{h. c. } \; .
\nonumber \\
\end{eqnarray}
In the Lagrangian (\ref{LHiggs}), $\Phi=\left(\underline{{\bf 2}},+1/2,0\right)$ is the usual SM doublet, $\Xi=\left(\underline{{\bf 1}},0,+2\right)$ is a scalar singlet needed for $A'$ acquires mass, $\left\{ \, \mu_{\Xi} \, , \, \mu_{\Phi} \, , \, \lambda_{\Xi} \, , \, \lambda_{\Phi} \, , \, \lambda \, \right\}$ are real parameters,
$\left\{ \, f_{ij}^{(\ell)} \, , \, f_{ij}^{(u)} \, , \, f_{ij}^{(d)} \, , \, f_{ij}^{(\nu)}
\, \right\}$
are Yukawa (complex) coupling parameters needed for the fermions to acquire non-trivial masses,
and $M_{\zeta}$ is a Dirac mass for the $\zeta$-fermion. In general, these Yukawa couplings set non-diagonal $3 \times 3$-matrices,
and, as usually, the $\widetilde{\Phi}$-field is defined as $\widetilde{\Phi}=i \, \sigma_{2} \, \Phi^{\ast}$.
The Majorana mass term for the Right-Handed Neutrinos (RHNs) was omitted because it should rather be introduced in the context of a
$Z'$-model for which the SSB mechanism takes place at the TEV scale. We here consider a SSB at the lower (GeV) energy scale.


%
The covariant derivatives act on the scalar fields as below:
\begin{eqnarray}\label{DmuXiProto}
D_{\mu} \Phi &=& \left(\partial_{\mu}
+ i g \, A_{\mu}^{\, a} \, \frac{\sigma^{a}}{2}+ i \, Y_{\Phi} \, g_{y} \, Y_{\mu} \right) \Phi \; ,
\nonumber \\
D_{\mu} \Xi &=& \left( \phantom{\frac{1}{2}} \!\!\!\! \partial_{\mu}
+i \, X_{\Xi}
\, g_{x} \, X_{\mu} \right) \Xi \, .
\end{eqnarray}
The Higgs potential in (\ref{LHiggs}) acquires two stable VEV scales, that we refer to as $v$ and $v_{x}$, so that the
the scalars fields have their corresponding VEVs, $\langle \Phi \rangle_{0} = \left( \, 0 \; \; v/\sqrt{2} \, \right)^{t}$
and $\langle \Xi \rangle_{0} = v_{x}/\sqrt{2}$, respectively. The scalars fields are parameterized
in the unitary gauge. Since $v=246 \, \mbox{GeV}$ is the VEV of SM, and $v_{x}$ generates the $A'$-boson mass,
in which we consider the condition $v_{x} \ll 246 \, \mbox{GeV}$ from now on.

Therefore, we are ready to obtain the masses,
the physical fields and the corresponding interactions of $A'$ with the fermions of the model.

\section{The masses and neutral currents}
After the SSB takes place, the Lagrangian for the neutral gauge fields can be cast in a matrix form as follows:
%
%
%
%
%
%
%
\begin{eqnarray}\label{LgaugeMatriz}
{\cal L}_{gauge}=-\frac{1}{4} \, K_{AB} \, F_{\mu\nu}^{ \; \; \; \; A} \, F^{\mu\nu \, B}
+ \frac{1}{2} \, M_{AB}^{\, 2} \, A_{\mu \, A} \, A^{\mu \, B}+J_{\mu \, A} \, A^{\mu \, A} \; ,
\end{eqnarray}
where the indices mean $A,B=\left\{ \, Y^{\mu} \, , \, A^{\mu 3} \, , \, X^{\mu} \, \right\}$ whereas $K_{AB}$ are the elements of the kinetic matrix
\begin{eqnarray}
K:=
\left(
\begin{array}{ccc}
1 & 0 & - \chi \\
0 & 1 & 0 \\
-\chi & 0 & 1 \\
\end{array}
\right) \; .
\end{eqnarray}
The mass matrix, $M_{AB}^{\, 2}$, is given by
\begin{eqnarray}
M^{2}=
\left(
\begin{array}{ccc}
g^{2}v^2/4 & -gg_{y} v^2/4 & 0
\\
\\
-gg_{y}v^2/4 & g_{y}^{\, 2} v^2/4 & 0
\\
\\
0 & 0 & 4 \, g_{x}^{2} \, v_{x}^{2} \\
\end{array}
\right)
\; ,
\end{eqnarray}
and the current $J_{A}^{\; \; \mu}$ is defined by
\begin{eqnarray}
J_{A}^{\; \; \mu}=\sum_{i} Q_{A}^{i} \, \bar{\psi}_{i} \gamma^{\mu} \psi_{i}
+i \, Q_{A}^{\Phi} \, \, \Phi^{\dagger} \, \overleftrightarrow{\partial}^{\mu} \, \Phi
+i \, Q_{A}^{\Xi} \, \, \Xi^{\dagger} \, \overleftrightarrow{\partial}^{\mu} \, \Xi \; ,
\end{eqnarray}
where the $i$ fermion index runs $i=\left\{ \, u_{L} \, , \, d_{L} \, , \, \nu_{L} \, , \, e_{L} \, , \, u_{R} \, , \, d_{R} \, , \, N_{R} \, , \, e_{R} \, , \, \zeta_{L} \, , \, \zeta_{R} \, \right\}$.

The diagonalization of the kinetic term is carried out by defining the sort of dreibein
$K_{AB}=e_{\; \; \; A}^{a} \, e_{\; \; \; B}^{a}$, where the fields change as
$A_{\mu}^{\, \, \, a}=e_{\; \; \; A}^{a} \, A_{\mu}^{\, \, \, A}$, and we obtain the diagonal term
as $K_{AB} \, F_{\mu\nu}^{\; \; \; \; A}F^{\mu\nu \, B}=F_{\mu\nu}^{\; \; \; \; a} \, F^{\mu\nu \, a}$.
The dreibein corresponds to the matrix
\begin{eqnarray}
e_{\; \; \; A}^{a}=
\left(
\begin{array}{ccc}
1 & 0 & -\chi \\
0 & 1 & 0 \\
0 & 0 & \sqrt{1-\chi^2} \\
\end{array}
\right) \; ,
\end{eqnarray}
and its inverse satisfies the condition $e_{A}^{\; \; \; a} \, e^{B}_{\; \; \; a}=\delta_{A}^{\; \; \; B}$, or $e_{a}^{\; \; \; A} \, e_{A}^{\; \; \; b}=\delta_{a}^{\; \; \; b}$. The mass matrix in the $(a,b)$-index basis reads as given below:
\begin{eqnarray}
M_{ab}^{\, 2}=e_{\; \; \; a}^{A} \, M_{AB}^{\, 2} \, e_{\; \; \; b}^{B}
=\left(
\begin{array}{ccc}
\frac{g^2 v^2}{4}+\frac{4 \chi ^2 g_x^2 v_x^2}{1-\chi ^2} & - \frac{g g_{y} v^2}{4} & \frac{4 \chi  g_x^2 v_x^2}{1-\chi ^2}
\\
\\
-\frac{g g_{y} v^2}{4} & \frac{g_{y}^{\, 2}v^2}{4} & 0
\\
\\
\frac{4 \chi  g_x^2 v_x^2}{1-\chi ^2} & 0 & \frac{4 g_x^2 v_x^2}{1-\chi ^2} \\
\end{array}
\right) \; .
\end{eqnarray}
%
This mass matrix can be diagonalized by the orthogonal transformation $A_{\mu}^{\; \; a}=R^{a}_{\; \; i} \, A_{\mu}^{\; \; i}$, where
$R^{a}_{\; \; i}$ are the elements of the diagonalizing matrix
\begin{eqnarray}
R(\theta,\theta_{W})=
\left(
\begin{array}{ccc}
\cos\theta_{W} & -\sin\theta_{W} & 0 \\
\sin\theta_{W} & \cos\theta_{W} & 0 \\
0 & 0 & 1 \\
\end{array}
\right)
\left(
\begin{array}{ccc}
1 & 0 & 0 \\
0 & \cos\theta & -\sin\theta \\
0 & \sin\theta & \cos\theta \\
\end{array}
\right) \; ,
\end{eqnarray}
where $\theta$ is the $Z$-boson/$A'$-gauge boson mixing angle, and $\theta_{W}$ is the usual Weinberg angle.
The index $i$ in $A_{\mu}^{\, i}$ refers to the gauge fields in the mass basis (fully diagonal mass matrix), $A_{\mu}^{i}=\left( \, A_{\mu} \, , \, Z_{\mu} \, , \, A'_{\mu}  \, \right)$. Thereby, the diagonal mass matrix can be read as below:
\begin{eqnarray}\label{MD}
M_{D}^{2}=R^{t} \, M^{2} \, R=
\left(
\begin{array}{ccc}
0 & 0 & 0 \\
0 & M_{Z}^{\, 2} & 0 \\
0 & 0 & M_{A'}^{\, 2} \\
\end{array}
\right) \; ,
\end{eqnarray}
whose eigenvalues, for the condition $v_{x} \ll v$, are given by
\begin{eqnarray}\label{MZMA}
M_{Z}\simeq \frac{v}{2} \, \sqrt{g^2+g_{y}^{\, 2}}
\hspace{0.3cm} \mbox{and} \hspace{0.3cm}
M_{A'} \simeq \frac{2g_{x}v_{x}}{\sqrt{1-\chi^2}} \, \sqrt{1+\frac{\chi^2 \, g_{y}^{\, 2}}{g^2+g_{y}^{\, 2}}} \; .
\end{eqnarray}
The first eigenvalue of (\ref{MD}) is identified as the photon mass, and the eigenvalue $M_{A'}$ we refer as the mass of
the $A'$-physical field.
%
%
%
%
The Weinberg angle is $\sin^{2}\theta_{W}=0.23$, such that the parametrization of $g$ and $g_{y}$ is like the one of the SM :
$g=e/\sin\theta_{W}=0.62$ and $g_{y}=e/\cos\theta_{W}=0.34$, and the $\theta$-angle is defined by
\begin{eqnarray}
\tan(2\theta)=\frac{2 \, \chi \, \sin\theta_{W}}{\sqrt{1-\chi^2}}\frac{M_{Z}^{2}}{M_{A'}^{2}-M_{Z}^{2}}\simeq
-\frac{2 \, \chi \, \sin\theta_{W}}{\sqrt{1-\chi^2}} \; ,
\end{eqnarray}
where we may assume $M_{Z} \gg M_{A'}$. By adopting these values, the $Z$-boson mass comes out the same as in the SM, {\it i. e.}, $M_{Z}=91 \, \mbox{GeV}$, and the $M_{A'}$ in (\ref{MZMA}) is the dark photon mass, that depends on the two free parameters of model, $g_{x}$ and $v_{x}$,
and we can fix it as $M_{A'}= 8$ GeV, or below. The $\theta$-angle depends on the $\chi$- kinetic parameter, the $g$ and $g_{y}$
coupling constants. If we take $\chi=10^{-3}$, we get the bound $|\theta|< 10^{-3}$, and the previous expressions can be simplified with
$\sin\theta\simeq 0$. When $\chi \rightarrow 0$, the $\theta$-angle goes to zero, and the transformation (\ref{transf}) reduces to the usual
diagonalization process of the SM.
%

Using the previous relations, the transformation to the basis of physical fields reads
\begin{eqnarray}\label{transf}
Y_{\mu} &=& \cos\theta_{W} \, A_{\mu}-\sin\theta_{W} \, Z_{\mu}
+\frac{\chi \, A_{\mu}^{\prime}}{\sqrt{1-\chi ^2}} \; ,
\nonumber \\
A_{\mu}^{3} &=& \sin\theta_{W} \, A_{\mu} + \cos\theta_{W} \, Z_{\mu} \; ,
\nonumber \\
X_{\mu} &=&
\frac{A_{\mu}^{\prime}}{\sqrt{1-\chi ^2}} \; .
\end{eqnarray}
In this basis, the interactions between the photon, $Z-$ and $A'$-bosons and any chiral fermion, $\psi_{i}$, of the model are cast below :
%
%
\begin{equation}
{\cal L}^{\, int} = e \, J_{em}^{\, \mu} \, A_{\mu}+ e \, J_{Z}^{\, \mu} \, Z_{\mu}+J_{A'}^{\, \mu} \, A_{\mu}^{\, \prime} \; ,
\end{equation}
where the currents are defined by
\begin{eqnarray}
J_{em}^{\, \mu} = \sum_{i} Q_{em}^{\, i} \, \bar{\psi}_{i} \, \gamma^{\mu} \, \psi_{i}
\hspace{0.3cm} , \hspace{0.3cm}
J_{Z}^{\, \mu}&=&\sum_{i} Q_{Z}^{\, i} \, \bar{\psi}_{i} \, \gamma^{\mu} \, \psi_{i}
\hspace{0.3cm} , \hspace{0.3cm}
J_{A'}^{\mu}=\sum_{i} Q_{A'}^{\, i} \, \bar{\psi}_{i} \, \gamma^{\mu} \, \psi_{i} \; ,
\end{eqnarray}
whose the generators $Q_{Z}^{\, i}$ and $Q_{A'}^{\, i}$ are given by
\begin{eqnarray}
Q_{Z}^{\, i}=\frac{I_{3L}^{\, i}-\sin^2\theta_{W} \, Q_{em}^{\, i}}{\sin\theta_{W}\cos\theta_{W}}
\hspace{0.3cm} , \hspace{0.3cm}
Q_{A'}^{\, i}=\chi \, \tilde{g}_{y} \, Y^{i}+ \tilde{g}_{x} \, X^{i} \; .
\end{eqnarray}
Notice that a millicharge $\chi \, \tilde{g}_{y}$ emerges in the generator $Q_{A'}^{\, i}$, and the $Q_{Z}^{\, i}$ generator remains like the one of the SM.
The $A'$-gauge boson couples to a linear combination of currents associated with the hypercharge and the $X$-charge.
For convenience, we split the interaction of the dark-photon with any $f$-fermion of SM and the $\zeta$-fermion according to the expression
%
%
%
%
%
\begin{equation}\label{LintX}
{\cal L}^{int}_{A'}=\bar{f} \, \, \slash{\!\!\!\!A}' \, \left(c_{fV}-c_{fA} \, \gamma_{5}\right) f + \tilde{g}_{x} \, \bar{\zeta} \, \, \slash{\!\!\!\!A}' \, \zeta \; ,
\end{equation}
where $\tilde{g}_{x}:=g_{x}/\sqrt{1-\chi^2}$, $\tilde{g}_{y}:=g_{y}/\sqrt{1-\chi^2}\simeq 0.34$,
and the coefficients $c_{fV}$ and $c_{fA}$ are defined by
\begin{eqnarray}\label{cVcA}
c_{fV}=\frac{1}{2} \, \chi \, \tilde{g}_{y} \left( Y_{\,f_{L}}+Y_{\,f_{R}} \right)+\tilde{g}_{x} \, X_{f}
\hspace{0.5cm} \mbox{and} \hspace{0.5cm}
c_{fA}=\frac{1}{2} \, \chi \, \tilde{g}_{y} \left( Y_{\,f_{L}}-Y_{\,f_{R}} \right) \; .
\end{eqnarray}
The $A'$-$f$ interaction is not CP-invariant if the $\chi$ mixing kinetic parameter is $\chi \neq 0$.
%

We have obtained the physical fields, the mass spectrum of the model, and the interactions of a dark fermion candidate with the $A'$-gauge boson.
In next Section, we shall investigate its phenomenological consequences and the allowed parameter regions related to the DM sector.
%

%

%

%
%
%
\section{The relic density of dark matter and its phenomenology}
%
%

%
%

Two annihilation processes are important for the DM detection : (i) $ \zeta \, \bar{\zeta} \, \rightarrow \, A' \, \rightarrow \, f \, \bar{f}$
and (ii) $\zeta \, \bar{\zeta} \, \rightarrow \, A' \, A' $. In both cases, the gauge boson $A'$ works as a portal that
connects the SM $f$-fermion with the $\zeta$-fermion DM candidate. Before the phenomenological analysis, we introduce the cosmological conditions needed to stabilize of $\zeta$-fermion as a DM candidate. We consider $\zeta$ and $A'$ in the thermal equilibrium in the early Universe, whose conditions are given by
\begin{enumerate}
\item
%
$n_{\zeta} \, \sigma\left(\zeta \, \bar{\zeta} \to f \, \bar{f}\right) > T^2/M_{Pl}$ \; ,
%
\item
%
$n_{A'} \, \sigma(A' \, A' \to f \, \bar{f}) > T^2/M_{Pl}$ \; ,
%
\end{enumerate}
where $T \gg M_{\zeta},M_{A'}$, $n$ is the number density of $\zeta$ or $A'$, and the Planck constant is $H \sim T^{2}/M_{Pl}^{\, 2}$
with the Planck mass $M_{Pl}=1.22 \times 10^{19}$ GeV. Under these conditions, the ratio $T^{2}/M_{Pl}^{\, 2}$ is the Hubble constant.
Using $n \sim T^{3}$ for both cases, $\sigma\left(\zeta \, \bar{\zeta} \to f \, \bar{f}\right) \sim \tilde{g}_{x}^{2}\left(c_{fV}^{2}+c_{fA}^{2}\right)/T^{2}$ and $\sigma\left(A' \, A' \to f \, \bar{f}\right) \sim \left(c_{fV}^{2}+c_{fA}^{2}\right)^2/T^2$ , yields
\begin{eqnarray}
n_{\zeta} \, \sigma\left(\zeta \, \bar{\zeta} \to f \, \bar{f}\right) &=& \tilde{g}_{x}^{2}\left(c_{fV}^{2}+c_{fA}^{2}\right)T \; ,
\nonumber \\
n_{A'} \, \sigma\left(A' \, A' \to f \, \bar{f}\right) &=& \left(c_{fV}^{2}+c_{fA}^{2}\right)^2T \; .
\end{eqnarray}
The freeze-in temperature is defined by
\begin{eqnarray}
\left. n \sigma \right|_{T=T_{FI}}=H(T_{FI})\sim\frac{T_{FI}^{\; 2}}{M_{Pl}} \; ,
\end{eqnarray}
where $\zeta$ and $A'$ get the thermal equilibrium with the SM particles. Requiring $T_{FI} > M_{\zeta} \, , \, M_{A'}$,
we can set up the lower bounds :
%
%
$\tilde{g}_{x}^{2} \left(c_{fV}^2+c_{fA}^{2}\right)>M_{\zeta}/M_{Pl}$ and $\left(c_{fV}^2+c_{fA}^{2}\right)^2>M_{A'}/M_{Pl}$. Therefore,
$\tilde{g}_{x} \sqrt{c_{fV}^2+c_{fA}^{2} } > 10^{-9}$ satisfies the conditions for the masses $M_{\zeta}$ and $M_{A'}$ at the GeV-scale
order, or below. We are now ready to push forward the DM analysis by calculating the relic density associated with the $\zeta$-fermion.
%

%
%
%

\begin{figure}[tb]
\vspace{-5pt}
\centering
\includegraphics[width=0.78\textwidth]{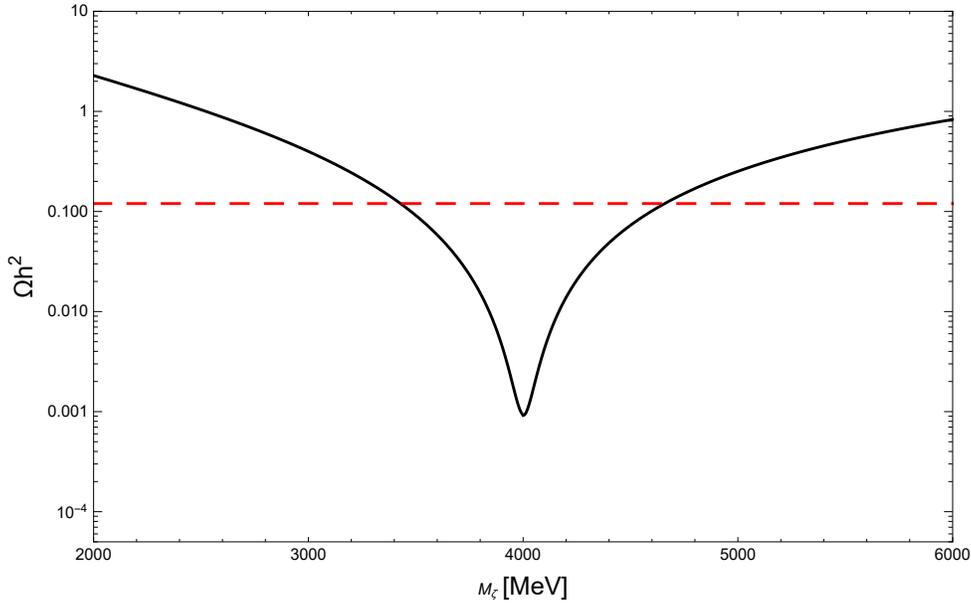}
\caption{A plot of DM relic density vs. $M_{\zeta}$ for $g_{x}=0.4$ and $M_{A'}\simeq 8$ GeV,
along with the observed relic density (dashed horizontal line).} \label{DMRD_sample}
\end{figure}

The DM relic density is the asymptotic solution to the Boltzmann equation \cite{review1,review2} :
\begin{eqnarray}\label{Omega}
\Omega_{DM} h^{2}
\simeq \frac{ 1.07 \times 10^{9} \, x_{f}}{\sqrt{g_{\ast}} \, M_{Pl} \, \langle v_{r}\sigma \rangle} \, ,
\end{eqnarray}
where $x_{f}$ is the ratio of DM mass by the freeze-out temperature of the DM particle
\begin{eqnarray}
x_{f}= \frac{M_{\zeta}}{T_{f}} \simeq \ln(X)-0.5\ln(\ln(X)) \; .
\end{eqnarray}
The dimensionless parameter, $X$, is defined by
\begin{eqnarray}
X = 0.038 \sqrt{\frac{g_{DM}}{g_{\ast}}} M_{Pl} \, M_{\zeta} \, \langle v_{r}  \sigma \rangle \; ,
\end{eqnarray}
where $g_{\ast}=106.75$ is the effective total number of degree of freedom for the particles in thermal equilibrium,
$g_{DM}=4$ is the number of degrees of freedom for the DM particle and
$\langle v_{r}  \sigma \rangle$ is the thermal average of the relative velocity $(v_{r})$ times the annihilation cross section $(\sigma)$ .
We approximate $\langle v_{r}  \sigma \rangle \simeq v_{r}  \sigma$ because the process is an $s$-wave annihilation.
We start off the analysis with the first annihilation process : $\zeta \, \bar{\zeta} \to A' \to f \, \bar{f}$.
Using known quantum field-theoretic rules, the annihilation cross section times the relative velocity at tree-level, in the non-relativistic limit, turns out to be
\cite{FarinaldoPRD2015}
\begin{widetext}
\begin{eqnarray}\label{vsigma}
v_{r} \sigma (\zeta \, \bar{\zeta} \to A' \to f \, \bar{f}) \simeq \frac{\tilde{g}_{x}^{\, 2}}{2\pi} \, N_{c}^{\, (f)} \,
\frac{2c_{fA}^{\, 2} (M_{\zeta}^{2}-m_{f}^{2})+c_{fV}^{2}( 2M_{\zeta}^{2}+m_{f}^{2}) }{(M_{A'}^{\, 2}-4M^2_\zeta)^2+M_{A'}^{\, 2} \, \Gamma^2_{A'}} \sqrt{1-\frac{m_f^2}{M_\zeta^2}} \; ,
\end{eqnarray}
\end{widetext}
%
%
where the total decay width of $A'$-boson is given by
%
\begin{eqnarray}\label{decayX}
\Gamma_{A'} &=& \frac{M_{A'}}{12\pi} \sum_{f} N_{c}^{\, (f)}
\sqrt{1-\frac{4m_{f}^{\, 2}}{M_{A'}^{\, 2}}}\left[ c_{fV}^{2}\left(1+\frac{2m_{f}^{\, 2}}{M_{A'}^{\, 2}} \right)
+c_{fA}^{2}\left(1-\frac{4m_{f}^{\, 2}}{M_{A'}^{\, 2}} \right) \right]\Theta(M_{A'}-2m_{f})
\nonumber \\
&&
\hspace{-0.5cm}
+ \frac{\tilde{g}_{x}^{\, 2} \, M_{A'}}{12\pi}
\sqrt{1-\frac{4M_{\zeta}^{\, 2}}{M_{A'}^{\, 2}}} \left(1+\frac{2M_{\zeta}^{\, 2}}{M_{A'}^{\, 2}} \right) \Theta(M_{A'}-2M_{\zeta}) \; ,
\end{eqnarray}
%
$m_{f}$ standing for the mass of SM fermions such that $m_{f} < M_{A'}/2$ and
$M_{\zeta} < M_{A'}/2$ for a Dark fermion candidate.
In (\ref{vsigma}) and (\ref{decayX}), $N_{c}^{\, (f)}$ is the color number of each fermion in the final state:
$N_{c}^{(Q)}=3$ for a quark, $N_{c}^{(\ell)}=1$ for a charged lepton, $N_{c}^{(\nu)}=1/2$ for a SM neutrino.
We fix the $A'$-mass at $M_{A'}=8.0$ GeV and $g_{x}=0.4$,
under these conditions, the DM relic density vs. $M_{\zeta}$ is plotted in figure (\ref{DMRD_sample}).

The observed DM relic abundance is described by the red line in the plot (\ref{DMRD_sample}),
whose value known in the literature is $\Omega h^2=0.12$ \cite{Planck2018}. The intersection
of the relic density curve with the red line yields the $\zeta$-mass at $M_{\zeta}=3.4$ GeV and $M_{\zeta}=4.6$ GeV.
If the decay $A' \, \rightarrow \, \zeta \, \bar{\zeta}$ is kinetically allowed, we may take
the mass $M_{\zeta}=3.4$ GeV in this process.

\begin{figure}[t]
\vspace{-5pt}
\centering
\includegraphics[width=0.78\textwidth]{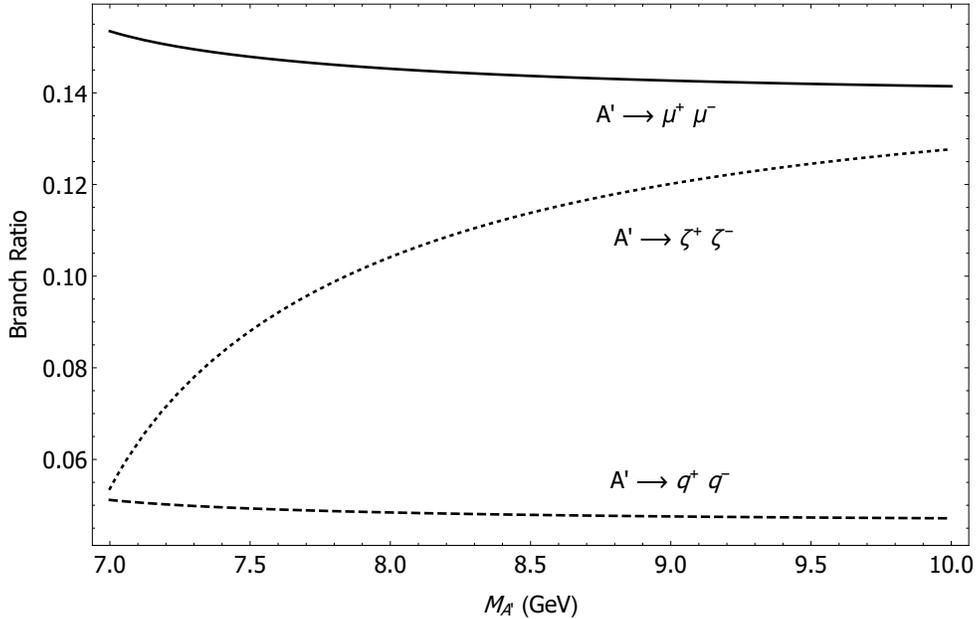}
\caption{
The branch ratio of the processes $A' \rightarrow \mu^{+} \, \mu^{-}$ (black line), $A' \rightarrow \bar{q} \, q$ (dashed line) and
$A' \rightarrow \zeta^{+} \, \zeta^{-}$ (dotted line) as function of $M_{A'}$ at the GeV scale, for $M_{\zeta}=3.4$ GeV and $g_{x}=0.4$.
}
\label{Br}
\end{figure}

The $A'$ phenomenology points out to the decay $A' \, \rightarrow \, \mu^{+} \, \mu^{-}$
as one of the processes in the final state of $e^{+} \, e^{-} \, \rightarrow \, \mu^{+} \, \mu^{-} \, A^{\prime}$.
The result (\ref{decayX}) reads the following decays
\begin{eqnarray}
\Gamma(A' \rightarrow \mu^{+}\mu^{-})\simeq\frac{\tilde{g}_{x}^{2} \, M_{A'}}{12\pi}\simeq 34 \, \mbox{MeV} \; ,
\end{eqnarray}
where we use $M_{A'} \gg m_{\mu}$, and the same result holds for $A' \, \rightarrow \, e^{+} \, e^{-}$.
The branch ratio of $A' \rightarrow \mu^{+}\mu^{-}$, whenever $M_{A'}=8.0$ GeV and $M_{\zeta}=3.4$ GeV, is given by
\begin{eqnarray}
\mbox{Br}(A' \rightarrow \mu^{+}\mu^{-})=\frac{\Gamma(A' \rightarrow \mu^{+}\mu^{-})}{\Gamma_{A'}} \simeq 0.15 \; .
\end{eqnarray}
The process $e^{+} \, e^{-} \, \rightarrow \, \gamma \, A'$ is searched for by the BABAR detector \cite{BaBarPRL2017} followed
by invisible decays of $A'$ in a 53 $\mbox{fb}^{-1}$ \cite{Babar2013,Babar20022013}. All the decay processes of $A'$ are contained
in the result (\ref{decayX}), if the process is kinetically allowed. The invisible sector is set by the decay $A' \, \rightarrow \, \zeta \, \bar{\zeta}$,
in which, if we use $M_{\zeta}=3.4$ GeV, the branch ratio of $A' \, \rightarrow \, \zeta \, \bar{\zeta}$ comes out given by
\begin{eqnarray}
\mbox{Br}(A' \, \rightarrow \, \zeta \, \bar{\zeta} ) \simeq 0.10 \; .
\end{eqnarray}
The BABAR experiment assumes that the single $A'$ exists in the range $0< M_{A'} \leq 8$ GeV; there, we plot the branch ratio
of the processes $A' \rightarrow \mu^{+} \, \mu^{-}$ and $A' \rightarrow \zeta \, \bar{\zeta}$ as function of $M_{A'}$ in figure (\ref{Br}).
The $A'$-decay width into quarks is kinetically allowed for $q=\left\{ \, u \, , \, d \, , \, s \, , \, c \, \right\}$.
The branch ratio of $A' \rightarrow \bar{u} \, u$ (and also is of $A' \rightarrow \bar{d} \, d$) reads below:
\begin{eqnarray}
\mbox{Br}(A' \, \rightarrow \, \bar{u} \, u) \simeq 0.048 \; .
\end{eqnarray}
%

%
%
%
%

\begin{figure}[t]
\vspace{-5pt}
\centering
\includegraphics[width=0.48\textwidth]{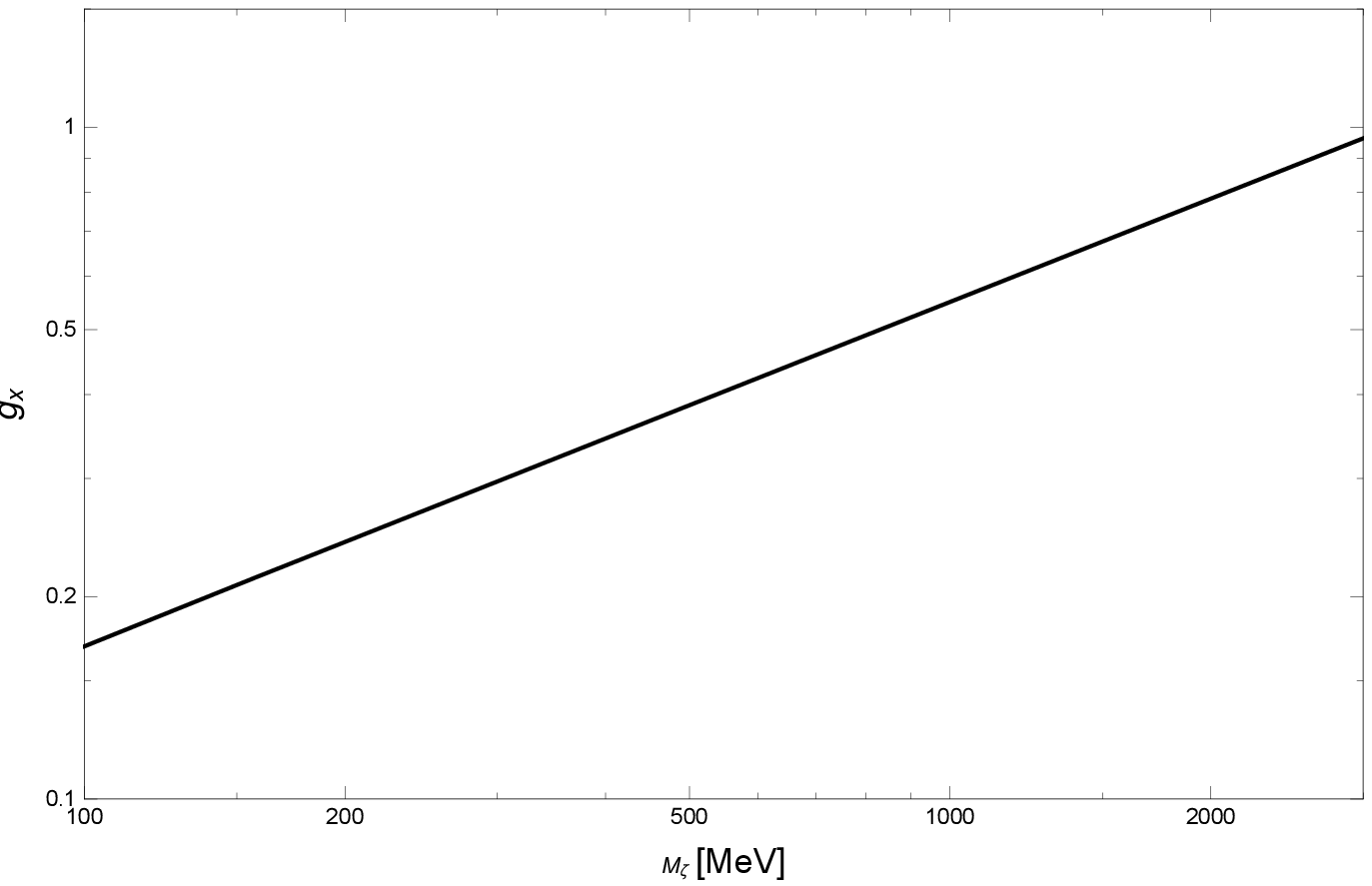}
\hspace{0.15cm}
\includegraphics[width=0.48\textwidth]{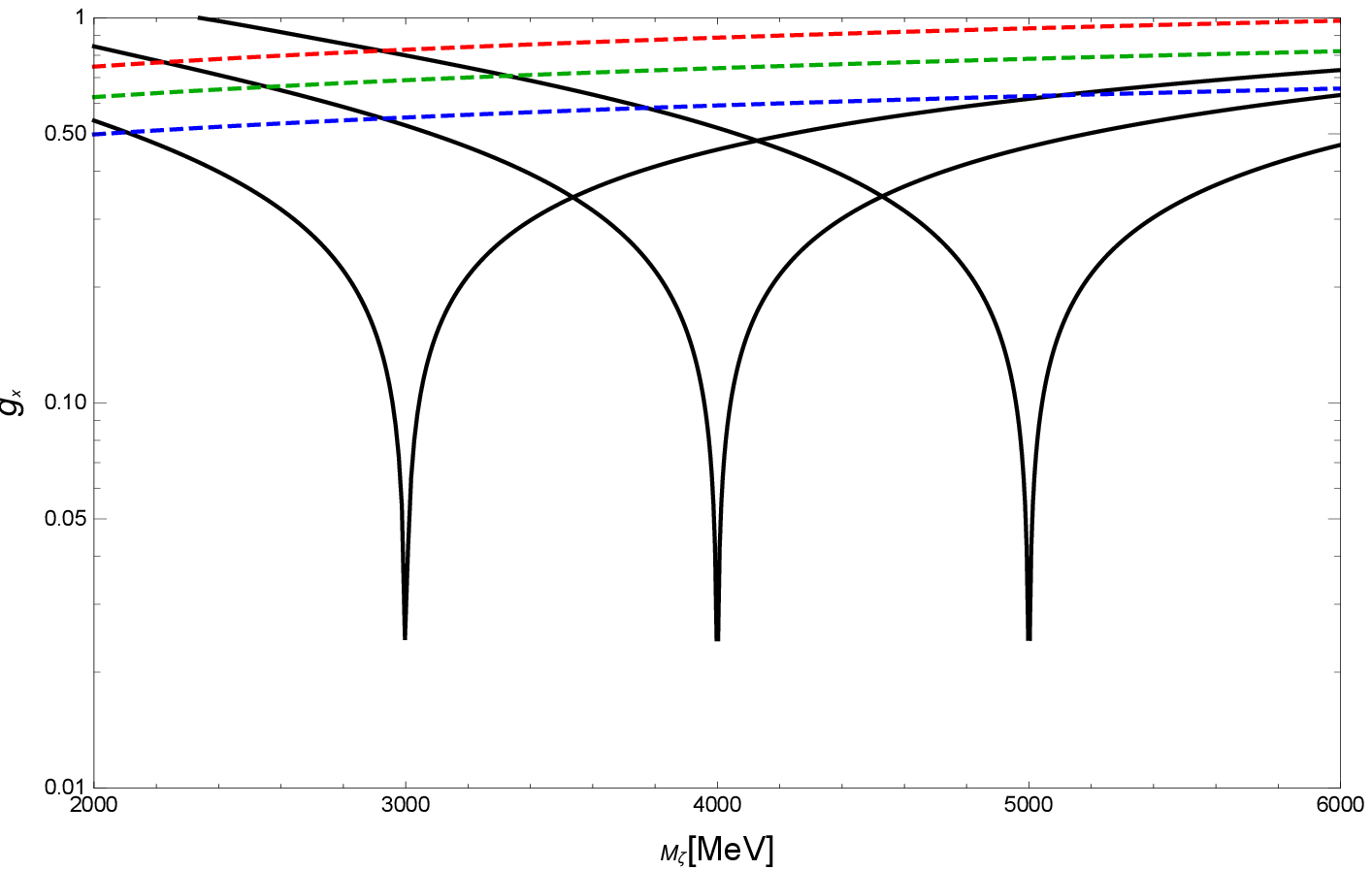}
\caption{
Left Panel : The $g_{x}$ as a function of $M_{\zeta}$ in which we have set $M_{A'}=M_{\zeta}/3$ for the second case $\overline{\zeta} \, \zeta \, \to \, A' \, A'$. Right Panel : The coupling constant $g_{x}$ versus $M_{\zeta}$ in the first annihilation process $\zeta \, \bar{\zeta} \to A' \to f \, \bar{f}$, for the masses of $M_{A'}=6.0$ GeV, $M_{A'}=8.0$ GeV and $M_{A'}=10.0$ GeV, respectively. The dashed lines (red, blue and green)
are the XENON1T experiment bounds. In both plots, the observed relic density is reproduced along the solid lines.
}
\label{AA}
\end{figure}

The second case is dominated by the annihilation process $\overline{\zeta} \, \zeta \, \to \, A' \, A'$, that is valid under the condition $M_{\zeta} > M_{A'}$.
In the non-relativistic limit, the annihilation cross section times the relative velocity can be read below
\begin{equation}
v_{r}\sigma(\overline{\zeta} \, \zeta \to A' \, A') \simeq \frac{\tilde{g}_{x}^{2}}{16\pi M_{\zeta}^{2}} \left(1-\frac{M_{A'}^{2}}{M_{\zeta}^{2}} \right)^{3/2}
\!\! \left(1-\frac{M_{A'}^{2}}{2 M_{\zeta}^{2}} \right)^{-2} .
\end{equation}
%
%
%
%
We depict $g_{x}$ versus $M_{\zeta}$ in figure (\ref{AA}) (Left panel), along which the observed DM relic abundance is reproduced.
In this analysis, we adopt $M_{A'}=M_{\zeta}/3$, as an example.
%
%
%

\section{The direct detection of dark matter}

The direct detection of DM is associated with the recent experiments XENON1T \cite{X1T} and LUX \cite{LUXLZ2017}.
The spin-independent cross section is the main equation for the elastic scattering of the DM particle with a nucleon (N),
via the process $\zeta \, N \to \zeta \, N$ mediated, in this model, by the $A'$-hidden gauge boson.
The expression for the spin-independent cross section is given by \cite{FarinaldoPRD2015,FarinaldoJHEP2015}
\begin{eqnarray}
\sigma_{SI}\simeq \frac{\mu^2_{\zeta N}}{\pi}\left[\frac{Zf_p+(A-Z)f_n}{A}\right]^{2} \; ,
\end{eqnarray}
where $\mu_{\zeta N}=M_{\zeta}M_{N}/(M_{\zeta}+M_{N})$ is the reduced mass of the system nucleon/dark fermion $\zeta$, $M_{N}$ is the nucleon mass,
$Z$ is the atomic number, $A$ is the atomic weight, and the parameters $f_{p}$ and $f_{n}$ are defined in terms of the coupling constants and $A'$-mass
\begin{eqnarray}
f_p=\frac{\tilde{g}_{x}}{M^2_{A'}} \left(2g_{u}+g_{d}\right) \; ,
\nonumber\\
f_n=\frac{\tilde{g}_{x}}{M^2_{A'}} \left(g_{u}+2g_{d}\right) \; .
\end{eqnarray}
We set $g_{u}=g_{d}=\tilde{g}_{x}/3$, so that $f_{p}=f_{n}$, and we obtain the relation
\begin{eqnarray}\label{SI}
\sigma_{SI}=\frac{\mu^2_{\zeta N}}{\pi} \, f_{p}^{\, 2}=\frac{\tilde{g}_{x}^{4}}{\pi} \, \frac{\mu_{\zeta N}^{\, 2}}{M_{A'}^{\, 4}} \; .
\end{eqnarray}
From the XENON1T experiment result, the upper bound for the nucleon scattering cross section is
\begin{eqnarray}\label{X1T}
\sigma_{SI} \leq 9.0 \times 10^{-11} \left(\frac{M_\zeta}{100 \; {\rm GeV}} \right) \, \mbox{pb} \; .
\end{eqnarray}
We plot these bounds in figure (\ref{AA}) (Right Panel with the dashed curves red, blue and green). Using the mass $M_{\zeta}=3.4$ GeV,
the upper bound for the cross section is $\sigma_{SI} \leq 3.0 \times 10^{-12}$ pb.

\section{The muon anomalous magnetic dipole moment}
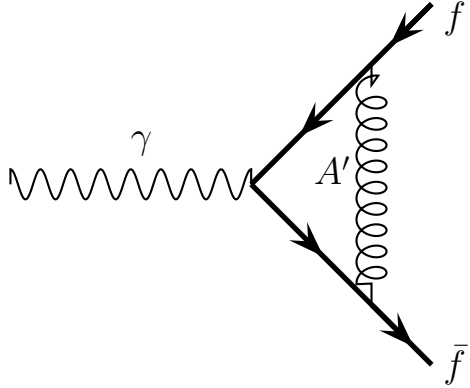
\begin{figure}[!h]
\begin{center}
\newpsobject{showgrid}{psgrid}{subgriddiv=1,griddots=10,gridlabels=6pt}
\begin{pspicture}(0,-0.5)(9,4)
\psset{arrowsize=0.2 2}
\psset{unit=0.8}
%
%
%
\pscoil[coilarm=0,coilaspect=0,coilwidth=0.5,coilheight=1.0,linecolor=black](1,2)(5,2)
\psline[linecolor=black,linewidth=0.7mm]{->}(8,5)(7.3,4.3)
\psline[linecolor=black,linewidth=0.7mm]{->}(8,5)(5.8,2.8)
\psline[linecolor=black,linewidth=0.7mm](6,3)(5,2)
\psline[linecolor=black,linewidth=0.7mm]{->}(5,2)(6.2,0.8)
\psline[linecolor=black,linewidth=0.7mm](6,1)(8,-1)
\psline[linecolor=black,linewidth=0.7mm]{->}(7,0)(7.7,-0.7)
\pscoil[coilarm=0.35,coilwidth=0.5,coilheight=1,linecolor=black](7,0)(7,4)
%
%
%
%
\put(3,2.6){\large$\gamma$}
%
%
\put(8.2,4.6){\large$f$}
\put(8.2,-1.2){\large$\bar{f}$}
\put(6.05,2){\large$A'$}
%
%
%
\end{pspicture}
%
%
\caption{\scshape{The one-loop correction of the $A'$-gauge propagator to the vertex photon-$f \bar{f}$.}}\label{VertexX}
\end{center}
\end{figure}

The anomalous magnetic moment of an $f$-fermion of the model is corrected by the diagram with photon vertex mediated by the
$A'$-propagator in the loop, as illustrated in the diagram (\ref{VertexX}).
%
%
%
Using usual rules of field theory, the correction is written in terms of the integral
\begin{eqnarray}\label{Deltaaf}
\Delta a_{f}(A')=\frac{N_{c}^{\, (f)}}{8\pi^2} \, \frac{m_{f}^{\, 2}}{M_{A'}^{2}} \int_{0}^{1} dz \, \,
\frac{c_{fV}^2 \, P_{V}(z)+c_{fA}^2 \, P_{A}(z) }{\left(1-z\right)\left(1-\lambda_{f}^{\, 2} \, z \right)+\lambda_{f}^{\, 2} \, z}
\; ,
\end{eqnarray}
where $z$ is a real parameter, $P_{V}(z)=2z^2(1-z)$, $P_{A}(z)=2z(1-z)(z-4)-4z^3 \, \lambda_{f}^{\, 2}$ and $\lambda_{f}=m_{f}/M_{A'}$, in which $m_{f}$ stands for the lepton or quark mass. To simplify (\ref{Deltaaf}), we use $\tilde{g}_{x} \gg \chi \, \tilde{g}_{y}$, such that we can approximate
$c_{fV}\simeq \tilde{g}_{x} \, X_{f}$, and to set $c_{fA}\simeq0$ when compared to $c_{fV}$.

\begin{figure}[t]
\vspace{-5pt}
\centering
\includegraphics[width=0.78\textwidth]{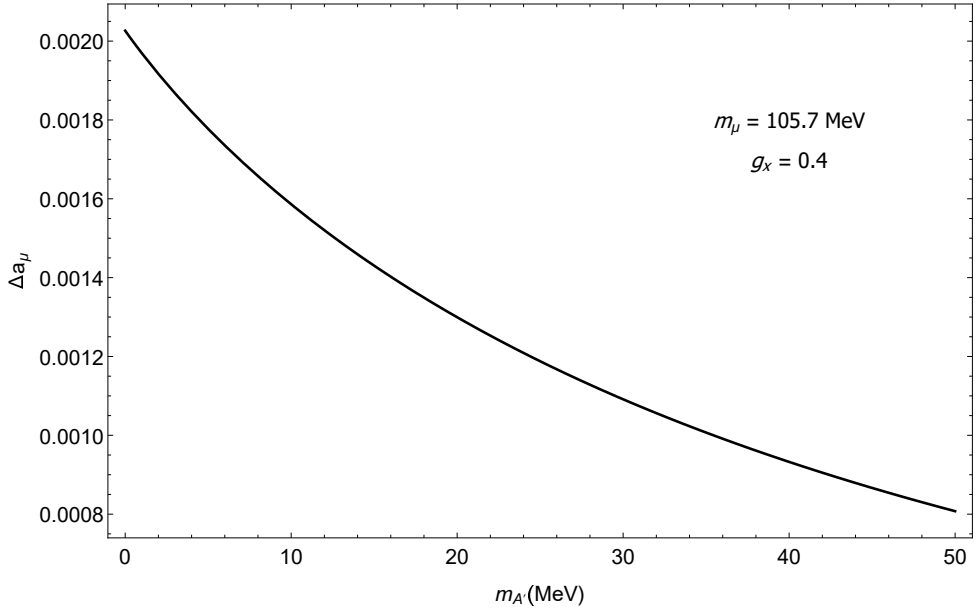}
\caption{
The correction to the muon's gyromagnetic ratio as function of $A'$-mass at the MeV scale, for $m_{\mu}=105.7$ MeV and $g_{x}=0.4$.
}
\label{muon}
\end{figure}

The muon magnetic moment is an interesting physical quantity that may reveal new physics beyond the SM. Using the mass $M_{A'}=8.0$ GeV, we get the ratio $\lambda_{\mu}=m_{\mu}/M_{A'}\simeq 0.013$ to yield the small result
\begin{eqnarray}
\Delta a_{\mu}(A')\simeq \frac{\tilde{g}_{x}^{2}}{12\pi^2} \, \frac{m_{f}^{\, 2}}{M_{A'}^{2}} \simeq 2.3 \times 10^{-7} \; .
\end{eqnarray}
If the $A'$-gauge boson is replaced by the $X$-boson, the mass $M_{X}=17$ MeV yields the ratio $\lambda_{\mu}\simeq 6.21$,
so that the muon anomalous moment receives a bigger contribution
\begin{eqnarray}
\Delta a_{\mu}(X)
\simeq 0.0014 \; ,
\end{eqnarray}
where we have used $\tilde{g}_{x}=0.4$. The plot of $\Delta a_{\mu}$ versus $A'$-mass at the MeV-scale
is illustrated in the figure (\ref{muon}).

\section{Concluding Comments}

We have devoted efforts to discuss a model with an extra $U(1)_{X}$-factor that may describe a possible scenario of a Particle Physics beyond the Standard Model (SM) at the GeV energy scale. The proposal is based on the gauge group $SU_{L}(2) \times U_{Y}(1) \times U(1)_{X}$, where the extra $U(1)_{X}$-factor introduces neutral gauge boson with a mass of $1-10$ GeV. The mass is generated upon a spontaneous symmetry breaking (SSB) mechanism that sets up an extra VEV scale $v_{x}$, beyond that VEV of $246 \, \mbox{GeV}$ associated with the usual Higgs field of SM.
The mechanism driven by the SSB pattern takes place at the VEV scale-$v_{x}$, where $v_{x} \ll 246 \, \mbox{GeV}$.
The model can be discussed in different scenarios: in connection with an $A'$-gauge boson, the $X$-boson or the paraphoton phenomenology, at $\mbox{GeV}$-scale or lower,
which can be related to DM particles. The $X$-boson case encompasses the recent description of a new boson needed to explain the excited
$8$-Beryllium nuclear decay. In this particular case, the mass of the $X$-boson is fixed at the mass scale of $M_{X}=17 \, \mbox{MeV}$.
We have calculated the relic density associated with a dark fermion candidate,
where the $A'$-gauge boson works as a DM portal through the two annihilation processes $\zeta \, \bar{\zeta} \, \rightarrow \, A' \, \rightarrow \, f \, \bar{f}$ and $\overline{\zeta} \, \zeta \, \to \, A' \, A'$. The observed relic density for the first case has yielded the mass around $M_{\zeta}=3.4$ GeV as a possible DM candidate, and we have worked out the decay modes of $A'$ with the branch ratio of $A' \rightarrow \mu^{+} \, \mu^{-}$, $A' \rightarrow q^{+} \, q^{-}$ and $A' \rightarrow \zeta^{+} \, \zeta^{-}$. The other important annihilation process is $\zeta \, \bar{\zeta} \, \rightarrow \, A' \, A' $. In this case, we have obtained the plot of the coupling constant $g_{x}$ vs. $M_{\zeta}$ for the observed relic density $\Omega h^2=0.12$. Next, we have briefly discussed the direct detection of DM through the elastic scattering process $\zeta \, N \, \rightarrow \, \zeta \, N$ and we have compared the parameters with the XENON1T experiment. Finally, we have analyzed the result for the 1-loop correction to the muon magnetic dipole moment. For the $A'$-gauge boson with an $8$ GeV mass, we have got the tiny value $\Delta a_{\mu}=2.3 \times 10^{-7}$; on the other hand, the $X$-boson model with $17 \, \mbox{MeV}$ mass yields $\Delta a_{\mu} \sim 0.0014$. This may be seen as an meaningful signal of a new physics beyond the SM.
\section*{Acknowledgments}
This work has been supported by the Conselho Nacional de Desenvolvimento Cient\'ifico e Tecnol\'ogico (CNPq) under grant
313467/2018-8 (GM).
\end{document}